\pdfoutput=1
\documentclass[sigconf]{acmart}

\copyrightyear{2024}
\acmYear{2024}
\setcopyright{rightsretained}
\acmConference[SIGIR '24]{Proceedings of the 47th International ACM SIGIR Conference on Research and Development in Information Retrieval}{July 14--18, 2024}{Washington, DC, USA} \acmBooktitle{Proceedings of the 47th International ACM SIGIR Conference on Research and Development in Information Retrieval (SIGIR '24), July 14--18, 2024, Washington, DC, USA} \acmDOI{10.1145/3626772.3657678} \acmISBN{979-8-4007-0431-4/24/07}


 
\usepackage{import,transparent}
\usepackage{nicefrac}
\usepackage{makecell}
\usepackage{algorithm}
\usepackage{algorithmic}
\usepackage{xspace}
\usepackage{pgfplots}
\usepackage{pgfplotstable}
\usepackage{booktabs}
\usepackage{multirow}
\usepackage{wrapfig}
\usepackage{paralist}
\usepackage{balance}
\usepackage[aboveskip=1ex, belowskip=-1ex]{caption}

\pgfplotsset{width=10cm,compat=1.9}
\usepackage{tikz}
\usetikzlibrary{shapes,shapes.multipart, shapes.geometric,backgrounds, shapes.arrows, arrows, patterns, positioning, calc}
\usepackage{graphicx}
\usepackage{xcolor}
\usepackage{pifont}

\usepackage[inline]{enumitem}

\usepackage{subcaption}

\usepackage{mdframed}

\newcommand{\eg}{e.\nolinebreak[4]\hspace{0.01em}\nolinebreak[4]g.\@\xspace}

\newcommand{\Reals}{{\mathbb R}}

\newcommand{\sysname}{CLIP-Branches\@\xspace}
\newcommand{\website}{\url{https://web.clip-branches.net/}}
\newcommand{\github}{\url{https://github.com/cluel01/clip-branches}}

\newcommand{\norm}[1]{\left\lVert#1\right\rVert}

\makeatletter
\gdef\@copyrightpermission{
  \begin{minipage}{0.3\columnwidth}
   \href{https://creativecommons.org/licenses/by-nc/4.0/}{\includegraphics[width=0.90\textwidth]{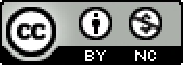}}
  \end{minipage}\hfill
  \begin{minipage}{0.7\columnwidth}
   \href{https://creativecommons.org/licenses/by-nc/4.0/}{This work is licensed under a Creative Commons Attribution-NonCommercial International 4.0 License.}
  \end{minipage}
  \vspace{5pt}
}
\makeatother
\begin{document}
%\title{Fast Search-By-Classification for Large-Scale Databases}
\title{\emph{\sysname}: Interactive Fine-Tuning for Text-Image Retrieval}

%%
%% The "author" command and its associated commands are used to define the authors and their affiliations.
\author{Christian Lülf}
\affiliation{%
  \institution{University of Münster}
  %\streetaddress{P.O. Box 1212}
  \city{Münster}
  %\state{Ohio}
  \country{Germany}
  \postcode{48153}
}
\email{christian.luelf@uni-muenster.de}

\author{Denis Mayr Lima Martins}
\affiliation{%
  \institution{Independent Researcher}
  %\streetaddress{P.O. Box 1212}
  \city{Campinas}
  %\state{Ohio}
  \country{Brazil}
  \postcode{48153}}
% \email{denis.martins@uni-muenster.de}
\email{denismartins@acm.org}
% \authornote{Work was primarily performed while the author was at the University of Münster.}

\author{Marcos Antonio Vaz Salles}
\affiliation{%
  \institution{Independent Researcher}
  %\streetaddress{P.O. Box 1212}
  \city{Cascais}
  %\state{Ohio}
  \country{Portugal}
}
\email{msalles@acm.org}
%\authornote{Work was primarily performed while the author was at the University of Copenhagen.}

\author{Yongluan Zhou}
\affiliation{%
  \institution{University of Copenhagen}
  %\streetaddress{P.O. Box 1212}
  \city{Copenhagen}
  %\state{Ohio}
  \country{Denmark}
}
\email{zhou@di.ku.dk}

\author{Fabian Gieseke}
%  \affiliation{%
%   \institution{University of Copenhagen}
%   \city{Copenhagen}
%   \country{Denmark}
% }
\affiliation{%
  \institution{University of Münster}
  \city{Münster}
  \country{Germany}
  \postcode{48153}}
\email{fabian.gieseke@uni-muenster.de}

%%
%% The abstract is a short summary of the work to be presented in the
%% article.
\begin{abstract}
The advent of text-image models, most notably CLIP, has significantly transformed the landscape of information retrieval. These models enable the fusion of various modalities, such as text and images. 
One significant outcome of CLIP is its capability to allow users to search for images using text as a query, as well as vice versa. This is achieved via a joint embedding of images and text data that can, for instance, be used to search for similar items. 
Despite efficient query processing techniques such as approximate nearest neighbor search, the results may lack precision and completeness. 
% While these queries can be efficiently implemented using well-known concepts, such as approximate nearest neighbor search, the results for a query might still be suboptimal, both from a precision and completeness perspective.
%Despite their remarkable zero-shot capabilities, these models often exhibit limitations in the accuracy and completeness of initial search results, necessitating fine-tuning to achieve optimal performance. 
% We introduce \textit{\sysname}, a novel text-image search engine that builds upon the CLIP architecture and improves search accuracy through an interactive fine-tuning phase. This phase allows users to refine queries using positive and negative examples, training a classifier to identify all positively classified instances
We introduce \textit{\sysname}, a novel text-image search engine built upon the CLIP architecture. Our approach enhances traditional text-image search engines by incorporating an interactive fine-tuning phase, which allows the user to further concretize the search query by iteratively defining positive and negative examples. Our framework involves training a classification model given the additional user feedback and essentially outputs all positively classified instances of the entire data catalog. 
By building upon recent techniques, this inference phase, however, is not implemented by scanning the entire data catalog, but by employing efficient index structures pre-built for the data.
%, but resorts to index structures to efficiently implement the inference phase. 
%. This phase involves training a machine learning model with user feedback on initial search results to improve the search outcomes. 
Our results show that the fine-tuned results can improve the initial search outputs in terms of relevance and accuracy while maintaining swift response times. 

\end{abstract}

\renewcommand{\authors}{Christian Lülf, Denis Mayr Lima Martins, Marcos Antonio Vaz Salles, Yong- luan Zhou, and Fabian Gieseke}
\maketitle
\section{Introduction}
\label{sec:introduction}
\begin{figure}[t]
    \centering    
    %\hspace{1.2cm}
    \scalebox{0.55}{%
        % \hspace{0.6cm}
        \input{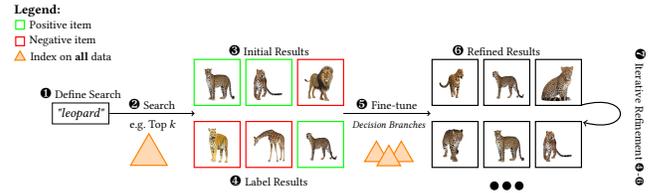}%
    }%
    \caption{Search process of \sysname. Traditional text-to-image search engines only consist of the steps \ding{182} to \ding{184} while \sysname adds a fine-tuning stage to refine the initial search results (Steps \ding{185}-\ding{188}). Index structures are employed during the search for faster execution.}
    \label{fig:search}
    %\vspace{-0.2cm}
\end{figure}

% \begin{itemize}[before=\color{red}]
%     \item Related work: The exploration of related work could be deepened to encompass additional advancements made on CLIP by others in the field.
%     \item extend the learning to rank part in the introduction -> mention relevance feedback, learning to rank in text
% \end{itemize}

Recent advances in deep learning have led to significant improvements in text-based image retrieval systems. Traditional methods, such as those employed by image platforms like Pinterest or Flickr, typically rely on nearest neighbor~(NN) search algorithms for data retrieval~\cite{pinterest}, where a set of similar items is returned for a given query text or image. These algorithms facilitate data retrieval through text and image embeddings, offering a rapid and efficient solution for handling extensive data sets through the use of appropriate index structures. The resulting answer set, however, might be suboptimal in the sense that desired objects of the data catalog might be missing or that objects not matching the query are given in the answer set.
%There is a growing need for systems that not only efficiently retrieve data but also incorporate user feedback to refine search results, tailoring them more closely to the specific needs of the user --- a feature largely absent in current engines.
Information retrieval approaches such as learning-to-rank~\cite{learning-to-rank} could be used to rank the set of nearest neighbors returned for a given query, e.g., using relevance feedback to increase search precision.
%\red{Information retrieval approaches such as learning-to-rank~\cite{learning-to-rank} could be used to filter or rank the set of nearest neighbors returned for a given query, thus increasing search precision.}
%One way to increase the search precision from an information retrieval perspective is to filter or rank the set of nearest neighbors returned for a given query \red{with techniques such as learning-to-rank}~\cite{learning-to-rank}.
However, such fine-tuning steps only consider the results returned by an NN approach, and typically still lead to a significant number of false negatives and positives.  
%the set of nearest neighbors  One may naturally consider the approach of \textit{Learning-To-Rank}, and use it to implement the fine-tuning of text-image retrieval results . 
%However, this approach is limited as it only re-evaluates the initially retrieved results without considering the broader data catalog. 
%Consequently, this limitation put us at risk of overlooking potentially superior results that were not identified in the initial NN search.

The development of text-image models, particularly %OpenAI’s 
CLIP~\cite{clip}, has substantially impacted the realm of text-based image retrieval. CLIP yields a joint embedding space for both text and image inputs and has given rise to various text-image search engines~\cite{beaumont-2022-clip-retrieval,clip-as-a-service}. 
%CLIP's strength lies in its training on a vast, general-purpose text-image pair data set.
One significant outcome of CLIP is, among other things,
its capability to enable users to search for images using text as a
query, as well as vice versa, which is achieved via the joint embedding space. As for the retrieval systems mentioned above, search queries can be implemented using (efficient) nearest neighbor search techniques. 
Nonetheless, searching for the nearest neighbors (e.g., top k) still often falls short in precision and relevance, i.e., the answer set for a query might contain examples not matching the query and might not contain all relevant examples. This is illustrated in Figure~\ref{fig:search} (Steps \ding{182}-\ding{184}). Here, the (text) query ``leopard'' yields an answer set of images. However, some images do not fit well to the query, and many leopard images might still be missing. This issue is further amplified by inherent limitations in text-image models like CLIP, which struggle to achieve fine-grained semantic understanding \cite{sigir2023}.

In this work, we present \sysname, a novel text-image search engine that extends traditional text-image search engines by incorporating relevance feedback through an iterative fine-tuning phase based on the concept of fast \textit{search-by-classification}~\cite{vldb}. Search-by-classification allows users to express their query intent by simply providing examples of what they seek (positive instances) and do not seek (negative instances). For example, given the search for ``leopard'' and the initial answer set, the user can further concretize the search intent by labeling some of the answer images as positive and some as negative, see Figure~\ref{fig:search} (Step \ding{185}).
These new labels can then be used to train a classification model, which, in turn, can be applied to the entire data catalog to retrieve more positive examples, see Figure~\ref{fig:search} (Steps \ding{186}-\ding{188}).
%Note that this would typically involve scanning the entire data catalog, resulting in a high response time not suited for search engines.
Instead of scanning the entire data catalog, which would result in high response times,\footnote{Applying a decision tree to a large data catalog might easily take minutes or hours.} we resort to so-called \textit{decision branches}~\cite{vldb}, a recently proposed variant of decision trees, which, in conjunction with pre-built index structures, allows us to efficiently implement the model application phase, leading to response times in the range of seconds only. Note that unlike traditional NN-based search engines that essentially only return (re-ranked) nearest neighbors for a given query, \sysname operates on the entire (!) database and retrieves all instances in the data catalog that are classified as positive by the model. Hence, this kind of retrieval is particularly beneficial in scenarios where users require a more complete set of query-relevant instances, which are obtained via an iterative and interactive refinement process.

%\paragraph{Contributions.}
The main contributions of this work are as follows:
\begin{enumerate}
    \item We present \sysname, a novel text-image search engine that enhances traditional models by incorporating user feedback for fine-tuning search results.\footnote{A prototype is available at: \website}
    %(about eight terabytes of image data). %The results affirm our system's capability to synergize speed, precision, and efficiency, redefining the image retrieval system benchmark.
    \item Several adaptations are needed to combine CLIP with decision branches. Among other things, we resort to suitable quantization techniques~\cite{jegou,ge_2013} to reduce the storage footprint of the CLIP embeddings. This is achieved by extending the neural network architecture used for CLIP and by adapting some of its weights using suitable loss functions.
%     To reduce the storage overhead associated with the required index structures for fast search-by-classification, we also minimize the storage footprint of the CLIP embeddings. Prior research has focused
% on minimizing the footprint of these embeddings by employing means of quantization~\cite{jegou,ge_2013}. \sysname builds upon these efforts, showcasing how to integrate quantization within a search-by-classification framework without compromising search quality.
    \item We demonstrate the effectiveness of \sysname across various extensive data sets encompassing over 260 million image instances. We also provide the source code that allows other researchers to (re)use the search engine for their own data and use cases. The code repository is publicly available.\footnote{\github}
\end{enumerate}
% In the rest of the paper, we first overview the technical details of \sysname in Section~\ref{sec:approach}. Afterwards, in Section~\ref{sec:demo}, we demonstrate how users can interact with \sysname. Conclusions are drawn in Section~\ref{sec:conclusions}.

% These examples train a machine learning classifier that, in conjunction with pre-built index structures, quickly classifies and retrieves new positive instances in large data catalogs. This technique has already shown its effectiveness in practical applications \cite{rapidearth}.

% This fine-tuning process is driven by user interaction, where users select relevant or irrelevant images in the initial search results. Through this feedback, the search engine refines its understanding of the user's query intent, leading to more accurate results.

Search-by-classification has already proven effective for geospatial search tasks~\cite{rapidearth}. \sysname introduces this method to multimodal text-image retrieval, leveraging CLIP embeddings to merge text and image data within a unified feature space.

\section{Technical Overview}
\label{sec:approach}
%\sysname introduces advancements in the realm of text-image search engines. %By design, such search engines are tasked with efficiently retrieving relevant image instances from extensive catalogs in response to text queries.
A key aspect of the functionality of text-image search engines is the rapid response time, a critical requirement given that users typically expect quick access to search results. These search engines predominantly leverage pre-built index structures, such as FAISS~\cite{douze2024faiss}, to achieve fast search times. Such index structures are generally populated with feature embeddings of the images that have been extracted beforehand, e.g., using deep neural networks. %These embeddings encapsulate crucial characteristics of the images, thus significantly reducing the storage resources required, since the original 

% \subsection{Search-by-Classification}
\subsection{Search Workflow} %and Architecture}
A general workflow of a text-image search engine is depicted in Figure~\ref{fig:search}. Note that traditional search engines usually only cover Steps~\ding{182}-\ding{184}. \sysname extends on that setup by adding a search-by-classification stage (Steps \ding{185}-\ding{188})~\cite{vldb}. After the first results have been retrieved in Step~\ding{184}, users can refine their search intent by selecting positive (correctly identified images) and negative (incorrectly identified images) instances (Step~\ding{185}). These instances serve as a training set for the decision branches~\cite{vldb} mentioned above. This model learns to discriminate between positive and negative objects using their feature embeddings. Once trained, the model is essentially applied to the entire data catalog, classifying the data into positive and negative instances (Step~\ding{186}). This operation is supported by pre-built index structures to speed up the inference and achieves interactive response times~\cite{vldb}. The user then receives all objects predicted as positive ranked by the model (Step~\ding{187}). If the user is not yet satisfied with the results, the iterative fine-tuning phase allows them to continuously refine the search results, thus progressively enhancing the accuracy and relevance (Step~\ding{188}).

\subsection{CLIP Embeddings}
\begin{figure}[t]
    \centering    
    %\hspace{1.2cm}
    % \scalebox{1}{%
    %     % \hspace{0.6cm}
    %     \input{figures/feature_extraction}%
    % }%
    \input{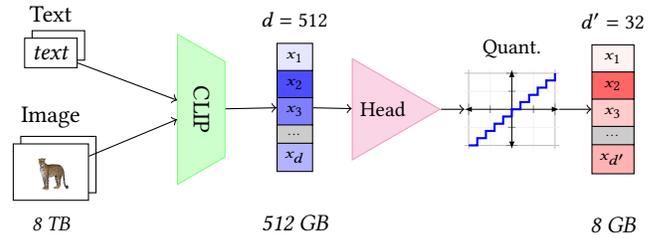}%
    \caption{Feature extraction for \sysname. The initial text and image data are transformed into 512-dimensional embeddings using CLIP. These embeddings are refined via a custom head module, followed by 8-bit quantization, resulting in a 32-dimensional 8-bit embedding. The reduction in data size is exemplified below using the LAION data set.
    }
    \label{fig:feat_extract}
\end{figure}

The extraction of appropriate embeddings is crucial for a search engine.
%'s precision and differs across various methods. 
Typically, the quality of embeddings improves with larger data sets or, in the case of feature extraction with deep neural networks, with larger model sizes. In the past, this has led to the common practice of adopting pre-trained large-scale models such as CLIP, which are subsequently fine-tuned for specific applications~\cite{pinterest2022}.

% \sysname operates on refined CLIP embeddings for text and images. 
\sysname resorts to CLIP embeddings for text and image data. As sketched in Figure~\ref{fig:feat_extract}, CLIP transforms %serves as the foundation for transforming 
raw image and text data into meaningful embeddings of dimensionality $d=512$. The extracted features typically capture many of the characteristics of the original data as the model was pre-trained on more than 400 million image-text pairs. However, significant storage is still required. We extend the CLIP model by attaching a custom head model (fully connected neural network) with $d'=32$ output neurons to reduce these storage requirements and to make the embeddings suitable for decision branches.
%To improve on this situation, we further refine the extracted feature vector by extending the original CLIP model with a custom model head. This head consists of extra fully connected layers attached to the end of the CLIP model, culminating in a final embedding layer. This final layer is a fully connected layer with an output dimensionality of only $d'=32$ that is used to extract the embeddings of our search engine. %, that resembles the operational dimensionality of our search engine. 
For the head module to yield meaningful results, it is essential to train the weights of the additional fully connected layers. To achieve this, we employ the MSCOCO Image Captioning (MSCOCO) data set~\cite{coco} with 82,612 training samples and 40,438 validation samples.\footnote{During this training phase, we freeze all the weights of the original CLIP model to exclusively fine-tune the newly added layers. The training was conducted over 100 epochs using a \textit{NVIDIA A100} GPU, under the same settings as the original CLIP model.}
To further reduce the storage consumption, we make use of a specialized regularization term, see Section \ref{sec:quant}. This term is specifically designed to favor uniformity in the spherical latent space of the final embedding and plays a crucial role in optimizing the features for the subsequent quantization process.

% To demonstrate the effectiveness of our search methodology, we conducted experiments across multiple image classification benchmark datasets. We compared when the $F_1$-score of our Decision Branches models exceeds the score of the $k$NN-classifier\footnote{$k$ was set to the number of existing positive instances in the test set.} considering the number of positive training samples (see Figure \ref{fig:search_performance}).

% \begin{itemize}
%     \item \red{table for feature accuracy vs original clip?}
% \end{itemize}

\subsection{Quantization}
\label{sec:quant}
An essential ingredient of \sysname is the reduction of the storage requirements of the embeddings. Similar to other approaches, we make use of quantization techniques to achieve this while not harming the expressiveness of the features. To this end, we adopt the so-called Kozachenko-Leononenko~(KoLeo) regularization initially introduced by~\citet{SablayrollesDSJ19} and later applied in the context of the well-known \textit{DINOv2} model~\cite{dinov2}. The regularization encourages the feature vectors to be spread out uniformly across the spherical feature space. This uniform distribution across the embedding space enhances the effectiveness of quantization. 
Given a batch of $n$ embeddings $x_1,x_2,\ldots,x_n \in \Reals^{d'}$ in a $d'$-dimensional space, KoLeo regularization is defined as 
\begin{equation}
\mathcal{L}_{KoLeo} = - \frac{1}{n} \sum^{n}_{i=1} \log(\rho_{n,i}),    
\end{equation}
where $\rho_{n,i} = min_{j \neq i}\norm{x_i - x_j}$ is the minimal Euclidean distance between $x_i$ and any other point in the batch. 
% \red{(maybe add mathematical formula for the regularization?)}.
For the quantization itself, we use an 8-bit scalar quantization method by mapping each value to one of 256 integer values. In total, the embedding size is reduced by a factor of 64 (more precisely, from 512 values with single floating point precision to 32 values with a 8-bit representation).

% \vspace{-0.5cm}

\subsection{System Setup}
% Our search engine is composed of three microservices: \textit{Data}, \textit{Web} and \textit{Search} apps. The Data app manages the raw image data required for displaying search results. The Web app, developed in NodeJS, serves as the user interface for the search engine. The core component, the Search app, contains the actual search logic and index structures.
%The search service was so designed that it could be deployed on standard hardware even for large-scale datasets, as long as an SSD with enough storage for the index structures is attached. 
% The individual services communicate via REST API calls, allowing for independent deployment on different servers. 
Before deployment, \sysname undergoes some pre-processing. This involves transforming the image data into quantized CLIP feature embeddings (see Figure \ref{fig:feat_extract}). Through this process, we can significantly reduce the storage footprint of the raw image data. For instance, the largest data set, the LAION data set, was reduced from 8~TB of image data to just 8~GB of 8-bit feature embeddings. %, a reduction of more than three orders of magnitude.
We then construct the required index structures for the initial search and for subsequent refinement using the decision branch models. In total, these index structures sum up to 62~GB of additional storage.
% in the case of the LAION data set. 
% We claim that, with this setup, the search service is capable of being hosted on servers with commodity hardware, even when managing large-scale data sets.
% such as LAION.
%We argue that in this setup the search service can be hosted on regular servers with commodity hardware even for large-scale datasets as LAION. 
%yielding storage requirements that are still over two orders of magnitude smaller than the raw image data.
% \subsection{Datasets}
\begin{figure}[tb]
    \centering    
    %\hspace{1.2cm}
    % \scalebox{1}{%
    %     % \hspace{0.6cm}
    %     \input{figures/feature_extraction}%
    % }%
    \scalebox{0.685}{
    \begin{tikzpicture}
\begin{axis}[
    xbar,
      width=10cm, % Adjust width as needed
    height=6cm, % Adjust height as needed
    enlargelimits=0.15,
    legend style={at={(0.5,-0.25)},
      anchor=north,legend columns=-1},
    xlabel={Number of Positive Instances},
    symbolic y coords={MNIST,EuroSAT,Caltech101,Food101,CIFAR10}, 
    ytick=data,
    % yticklabel style={font=\bfseries}, % Make y-axis labels bold
    bar width=6pt,
    nodes near coords,
    nodes near coords align={horizontal},
    ]
\addlegendentry{\textit{DBranch Ensemble}}
\addplot %+[bar shift=-2pt]
    coordinates {(16,CIFAR10) (8,Caltech101) (3,EuroSAT) % Updated coordinates
        (10,Food101) (3,MNIST)};

\addlegendentry{\textit{DBranch}}
\addplot %+[bar shift=0.3pt]
    coordinates {(45,CIFAR10) (26,Caltech101) (5,EuroSAT) % Updated coordinates
        (30,Food101) (3,MNIST)};

\legend{\textit{DBranch Ensemble},\textit{DBranch}}
\node[, fill=white, anchor=south east, text width=1.5cm,xshift=-1mm,yshift=2mm] at (rel axis cs:1,0) {
    \textcolor{red}{Mean: 21.8}\\
    \textcolor{blue}{Mean: \xspace8.0}
};

\end{axis}
\end{tikzpicture}%
    }
    \caption{Number of positive training instances required for Decision Branches models to surpass NN search results among different image classification benchmark data sets.}
    \label{fig:search_performance}
\end{figure}
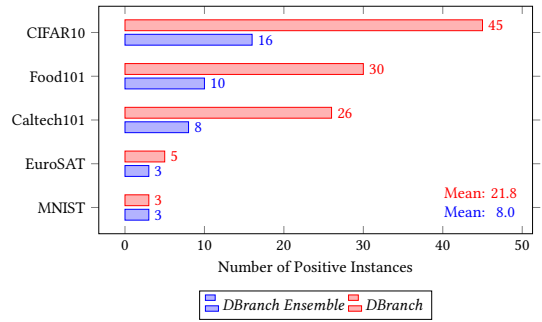

\begin{table}[tb]
\centering
\begin{tabular}{lcccc}
\hline
       & \multicolumn{3}{c}{MSCOCO}            & Accuracy \\ \cline{2-4}
                     Recall@ & 1 & 10 & 100 &                  \\
\hline
CLIP $(d=512)$  
&  -             & -               & -                & 0.474             \\            
% &  0.673             & 0.724               & 0.759                & 0.474             \\      

CLIP + PCA            & 0.683              & 0.757               & 0.802                & 0.144             \\ 
CLIP + Head     & 0.831              & 0.890               & 0.926               & 0.390             \\
% CLIP$_{32}$ (\textit{euc})      & 0.929              & 0.950               &  \textbf{0.967}               & 0.293             \\ 
CLIP + Head + KoLeo          & 0.953              &  0.970               & 0.985                & 0.404             \\ \hline
\end{tabular}
\vspace{1em}
\caption{Comparison of quantization methods applied to original CLIP features $(d = 512)$ for the MSCOCO data set, detailing recall scores at 1, 10, and 100 nearest neighbors post-quantization, alongside zero-shot classification accuracy based on the embeddings. Each quantization method reduces the feature dimensionality to $d = 32$.}
% \caption{Comparison of quantization methods applied on raw CLIP features ($d=512$) for the MSCOCO data set. It includes recall scores at 1, 10, and 100 nearest neighbors, reflecting the effectiveness in preserving the original neighbors after quantization. Additionally, it shows zero-shot classification accuracy scores on the embeddings. All quantization methods work on features of $d=32$.
%Additionally, it shows zero-shot classification accuracy scores based on the embeddings.}
\label{tab:quantization_comp}
\vspace{-0.1cm}
\end{table}
\section{Experimental Evaluation}
To assess the benefits of our fine-tuning process over standard text-image retrieval using NN search, we performed evaluations across various image classification data sets with our CLIP embeddings.
% To validate the effectiveness of our proposed fine-tuning stage in contrast to regular text-image retrieval based on NN search, we conducted experiments across multiple image classification data sets using our CLIP embeddings.
We compared when the $F_1$-score of our decision branches models exceeded the score of a classification based on NN search\footnote{All $k$ nearest neighbors found in the test set given a positive training instance are considered to be positive, while all other instances are considered to be negative. The number of $k$ is set to the true number of positives in the test set.  Note that such important information is not known in
practice and gives the NN search an advantage.} with an increasing number of positive training samples for our models (see Figure \ref{fig:search_performance}). In the context of our search engine, we could thereby show that on average 22 positive samples are sufficient to outperform initial search results (based on NN search) w.r.t. the $F_1$-score with single decision branches. In comparison, only 8 positive samples are required when using decision branch ensembles. Note that the user interface, see Section~\ref{sec:demo}, allows to select such instances quickly. 

To assess the impact of the quantization on the quality of the induced features, we resort to the \textit{Recall@$k$} metric as shown in  Table~\ref{tab:quantization_comp}. This involves comparing the quality of a set of nearest neighbors computed on one of the quantized features (such as CLIP + PCA followed by quantization), given the set of nearest neighbors computed on the same set of features without quantization (\eg CLIP + PCA). A Recall@$k$ value close to one indicates a good overlap, indicating minimal impact from quantization on the corresponding feature set. Note that we only report Recall@$k$ values for the reduced feature sets. Table ~\ref{tab:quantization_comp} illustrates that the features processed with our head and KoLeo regularization maintain neighborhood structures more effectively through the quantization step than those processed by traditional methods (e.g., CLIP + PCA) or those without KoLeo regularization (CLIP + Head). 

%, thus helping the user to start with reasonably-sized labeled sets.

%Concerning the quantization, we observed that our quantized CLIP embeddings , such as applying principal component analysis~(PCA) to the original features or omitting the use of the .
% We resort to the Recall@$k$ metric to compare the quality of the sets of nearest neighbors computed on the quantized features, given the set of nearest neighbors computed on the original features for $k \in \{1, 10, 100\}$. Here, a value of 1.0 corresponds to a perfect overlap, \ie, both sets are the same, whereas a value of 0.0 corresponds to no overlap.
% To assess the effectiveness of our approach, we employ the Recall@$k$ metric. This metric measures the proportion of the true nearest neighbors from the original feature vectors that are successfully identified within the top $k$ positions of the nearest neighbors list derived from the quantized feature vectors, for values of$k \in \{1, 10, 100\}$. A Recall@$k$ score of 1 indicates a perfect match, where all nearest neighbors found in the quantized embeddings correspond exactly to the true nearest neighbors from the original embeddings.
% of queries where the true nearest neighbor is included among the top $k$ candidates in the quantized embedding space (for $k \in \{1, 10, 100\}$). The results indicate a superior performance of the techniques introduced above.

Finally, the zero-shot classification results (Accuracy) shown in Table~\ref{tab:quantization_comp} indicate that employing the final CLIP + Head + KoLeo features yields the best accuracy, indicating that these embeddings retain a significant portion of the original information compared to the original CLIP features.
\begin{figure*}[t!]
    \centering
        \begin{subfigure}{0.33\textwidth}
        \fbox{\includegraphics[width=0.95\linewidth]{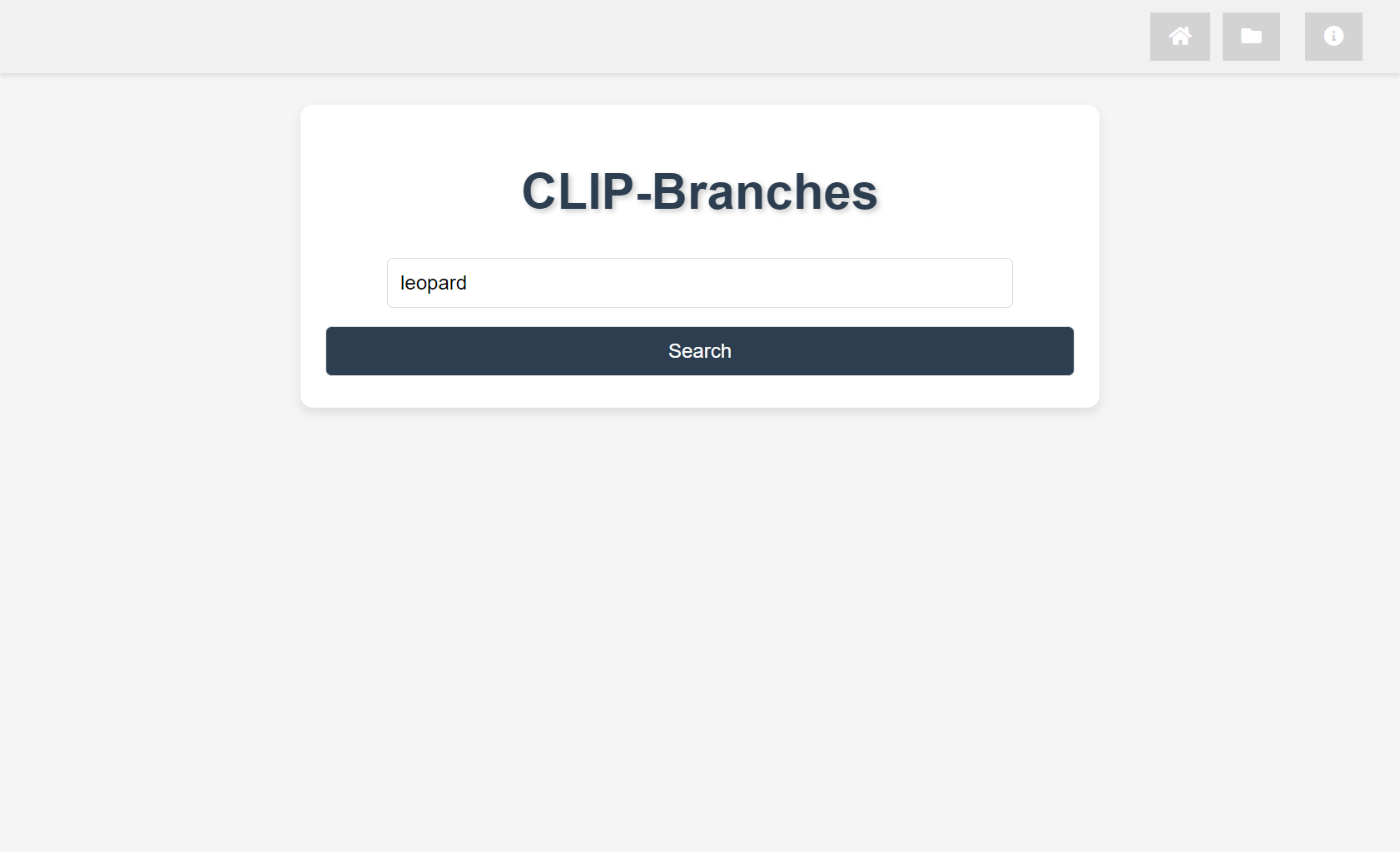}}
        \caption{Search Interface}
        \label{fig:ui}
    \end{subfigure}
    \begin{subfigure}{0.33\textwidth}
        \fbox{\includegraphics[width=0.95\linewidth]{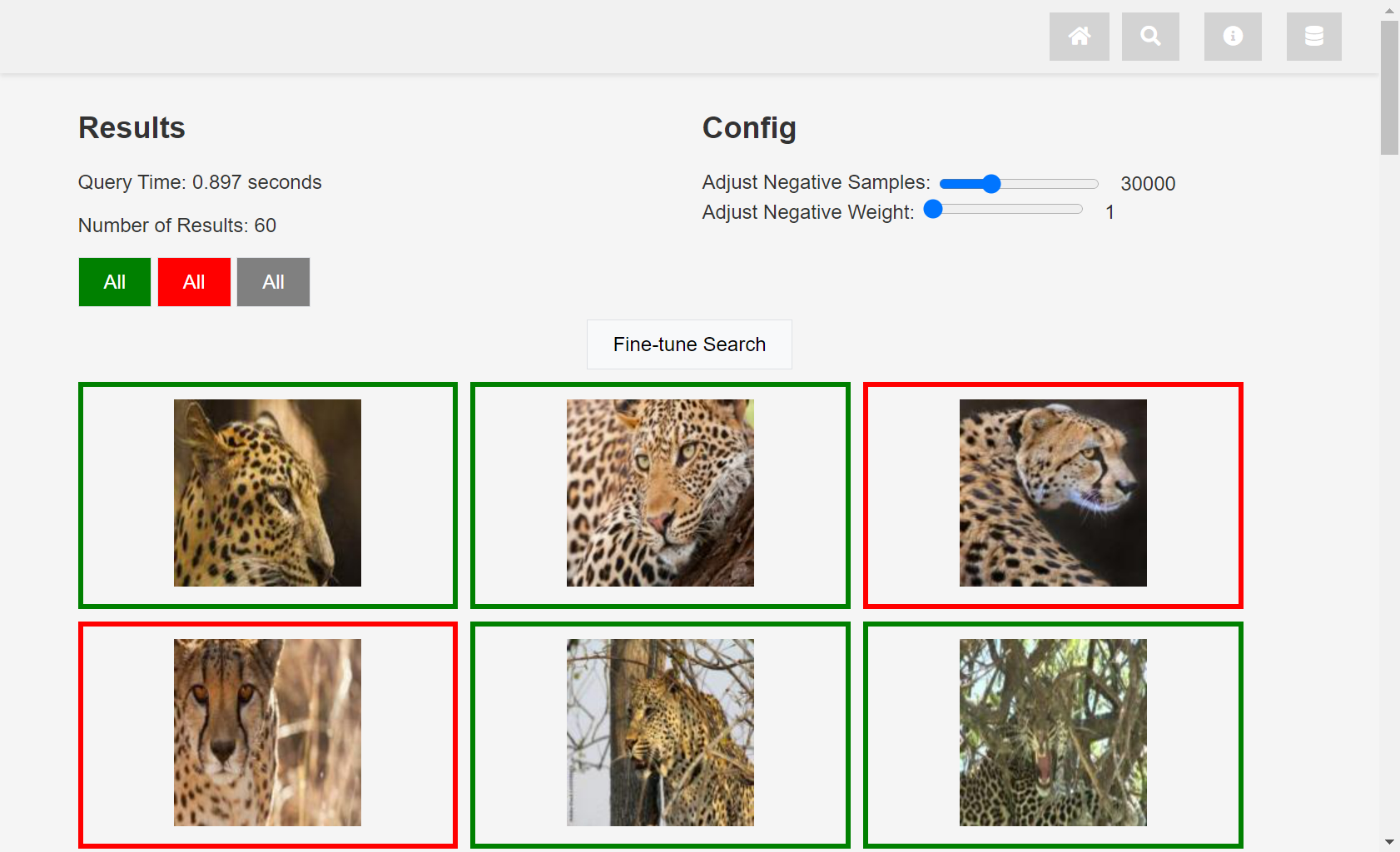}}
        \caption{First Top $k$-Results}
        \label{fig:topk}
    \end{subfigure}
        \begin{subfigure}{0.33\textwidth}
        \fbox{\includegraphics[width=0.95\linewidth]{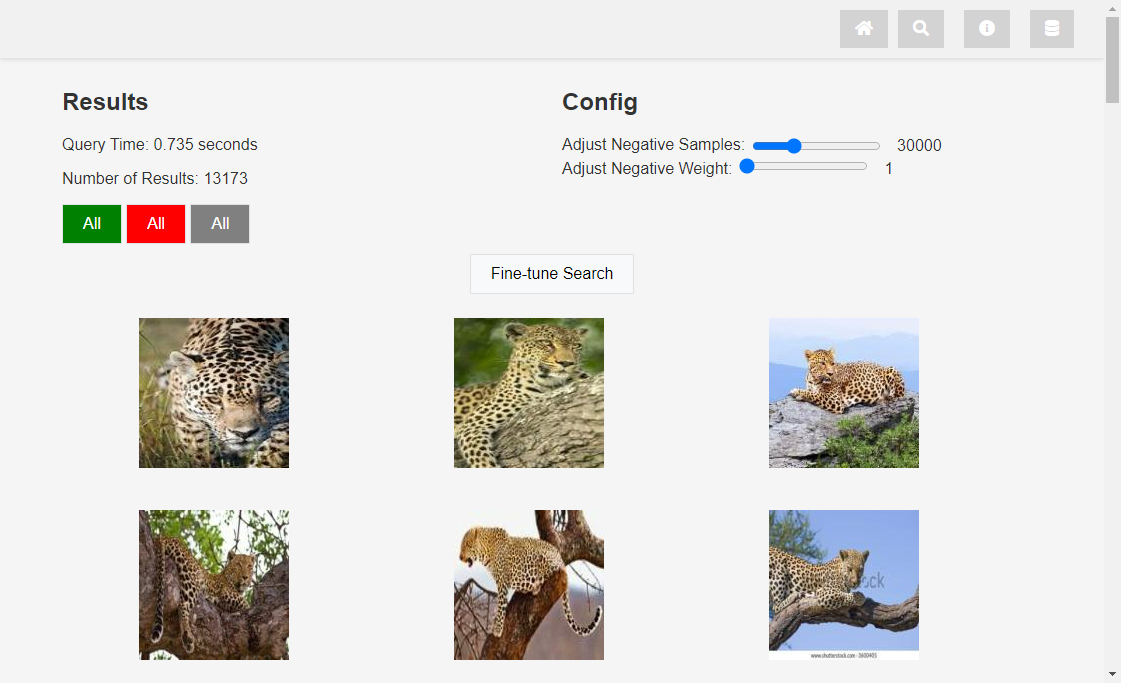}}
        \caption{Fine-tuned Results}
        \label{fig:finetune}
    \end{subfigure}
    \vspace{0.02cm}
    \caption{Demonstration of \sysname search workflow. The user initiates a search with a query string and receives top $k$-initial results. These results are then labeled as positive (green) or negative (red) based on user preference, guiding the fine-tuning of the search. The fine-tuned results reflect more completeness and higher accuracy.}
    \label{fig:demo}
    \vspace{-0.1cm}
\end{figure*}
\section{Demonstration}
\label{sec:demo}
In our demonstration, we show how users interact with our first prototype of \sysname. To effectively demonstrate its functionality, we have prepared multiple image data sets:
% , encompassing up to 260 million images:
\begin{itemize}
    \item \textbf{CIFAR10}: a subset of the CIFAR10 benchmark data set consisting of 50,000 images.
    \item \textbf{Shutterstock}: a data set collected from the Shutterstock platform\footnote{\url{https://shutterstock.com}} consisting of more than 14 million images~\cite{nguyen2022}.
    % \footnote{\url{https://shutterstock.com}}
    %, offering license-free stock footage with over 14 million images .
    \item \textbf{LAION}: a subset of the LAION-400M web-scraped data set with more than 260 million images \cite{schuhmann2022laion}.%\footnote{Disclaimer: a post-filtering of the images is based on appropriateness and duplicates is not done yet.}
\end{itemize}
The user can choose between various classification models for refining the search. In addition to decision branch models, we have also incorporated classical decision trees and random forests to highlight the differences in query time when using unoptimized models for the search:
% . The available models are: 
\begin{enumerate*}
    \item a single \textit{decision branch} model,
    \item a \textit{decision branch ensemble} comprising 25 individual models,
    \item a standard \textit{decision tree} and
    \item a \textit{random forest} model.
    %\item \textit{HyperNN}, a neural network-based variant of decision branches, adapted from \citet{LimaDCF2023EndToEndNeural}.%an extension of a neural network-based classification model that learns orthogonal decision boundaries .
\end{enumerate*}
%Next, we sketch the user journey through the \sysname interface, highlighting its features and capabilities. 
The general workflow and the interface are depicted in Figures~\ref{fig:search} and \ref{fig:demo}, respectively.

\begin{enumerate}
    \item \emph{Initiating the search:}  When entering the search engine, see Figure \ref{fig:ui}, users first encounter a text search interface. The top bar includes options to choose the data set (accessible via the "\textit{Folder}" icon) and a manual (via the ``\textit{Info}'' button). Users can input their search string into a dedicated text field and start the search via the respective button. Upon initiating the search, this string is converted into a CLIP feature representation for which the nearest neighbor embeddings are retrieved. To address the computational challenges associated with high-dimensional searches
    %\footnote{This problem is called the "curse of dimensionality" and refers to the degrading behavior of NN search in high-dimensional spaces \cite{curseofdimensionality}.}
    %(32-dimensional in our case), 
    we resort to an approximate nearest neighbor (ANN) search strategy supported by an index\footnote{We use an implementation of a $k$-d tree as ANN index \cite{Bentley1975}.} for the first search. This method significantly accelerates the retrieval process by approximating the closest neighbors rather than computing exact matches. %While this approach initially trades off accuracy for speed, we introduce a subsequent interactive fine-tuning phase to obtain better search results.
%While this approach trades off a degree of accuracy for speed, it's important to note that any potential discrepancies introduced at this stage are systematically addressed and refined in the subsequent fine-tuning phase.
% Since the exact search of nearest neighbors in 32-dimensional space can be quite slow due to the curse of dimensionality, we apply an approximate nearest neighbor search to speed up the search at the cost of the accuracy of the results. These sub-optimal results, however, can be corrected in the next step of the finetuning anyway.

    \item \emph{First results and interactive labeling:} After initiating the search, users are presented with an initial set of results, comprising up to 60 images that match their query, see Figure \ref{fig:topk}. Users can interact with these results by labeling images as positive (green frame), negative (red frame), or leaving them unlabeled (no frame). By using the ``\textit{All}'' buttons, users can simultaneously apply the corresponding label to all displayed images. In the top bar, users can select their preferred fine-tuning classification model via the ``\textit{Lens}''-button. 
In addition, users can tweak search parameters to refine the outcome. A key feature here is the adjustable slider for negative sample inclusion. By increasing the count of randomly sampled negative images, the precision of the model's results is enhanced, but at the expense of longer training times.\footnote{Random sampling assumes that negative instances far outnumber positive ones in the addressed tasks. When positive instances are falsely included in the negative sample, the machine learning models used are generally robust to this small amount of noise.}
% We advise a higher proportion of negative samples to balance the training dataset, considering the natural positive-negative ratio in typical datasets. 
Another adjustable parameter is the negative weight slider, which determines the influence of user-labeled negative samples relative to randomly chosen ones. 
% A high negative weight is recommended to ensure these self-labeled negatives are effectively excluded in subsequent searches. 
Once satisfied with the labeling and parameter configuration, the user can proceed to the next phase by clicking the ``\textit{Fine-tune Search}'' button. 
%This action launches the refined search, applying the user's feedback to generate more accurate results.

\item \emph{Fine-tune search:} The retrieval time depends on the selected model. In general, for quick searches, single decision branch models are well suited. Conversely, for higher accuracy, but with slightly longer wait times, decision branch ensemble models are preferable. For each run, search statistics are shown, see Figure \ref{fig:finetune}. 
%about the executed search, which are the search time and number of results, 
% Unlike traditional NN search engines, \sysname returns all positively classified instances, not just top-ranked results.
If the initial fine-tuned results are not satisfactory, users can iterate the fine-tuning process until the desired outcome is achieved.
% It has to be stressed that unlike nearest neighbor search engines, \textbf{all} instances that are classified by the model as positive are returned and not just the Top-$k$ results. Ideally, the fine-tuned results should already be of decent quality. If this, however, is not satisfying the needs of the user, he can fine-tune the search results over multiple iterations until the results are satisfying. At that point, the user can then browse through all the results and pick the instances he is interested in (e.g. export them).

% A notable integration feature is the ability to link the search engine with the database community. For those utilizing Decision Branches or HyperNN models, the system enables translation of these models into SQL range queries. These queries can be extracted via the "\textit{Database}" button in the top bar, facilitating the application of similar searches in other databases with equivalent feature embeddings.
\end{enumerate}
Users are invited to explore \sysname through the provided link: \website.
%\noindent\textbf{(1)  
% \todot{some text here? What is THE defining difference compared to existing search engines?}
%\noindent\textbf{(2) First Results and Interactive Labeling.} 

%\noindent\textbf{(3) Fine-tune Search.} 

%\footnote{\red{At the moment, only parts of the model are extracted (one box).}}

% Also, we implemented a feature to link the search engine to the database community. If the user worked with a Decision Branches model or HyperNN, the model can be translated into SQL range queries. These can be extracted via the "Database"-button in the top bar and then can be used to apply the same search\footnote{\red{At the moment, only parts of the model are extracted (one box).}} in other databases with the same feature embeddings.

\section{Conclusion}
\label{sec:conclusions}
We introduce \sysname, a text-image search engine that incorporates a novel refinement feature. 
Both the code for setting up the search engine as well as a video showcasing how users can interact with \sysname are available on GitHub.\footnote{\github}
% \red{link to a video: "The video should be hosted and the URL should be linked in both the CMT abstract and in the paper. Authors will be allowed to revise their video for the camera ready."}

\begin{acks}
    This research is supported by the Independent Research Fund Denmark
(grant number 9131-00110B). We also acknowledge support from NVIDIA for a hardware donation.
\end{acks}

\bibliographystyle{ACM-Reference-Format}
\balance
\bibliography{literature}

\end{document}